\input{aipcheck}

\documentclass[,final
  ]
  {aipproc}

\layoutstyle{8x11single}
\usepackage{amstext}
\usepackage{amsopn}
\usepackage{amsfonts}
\usepackage{amsmath}
\usepackage{amssymb}
\usepackage{color}
\usepackage{graphics}
\usepackage{graphpap}
\usepackage{amscd}
\usepackage{array}
\usepackage{graphicx}
\usepackage{amssymb,amsmath}
\def\noi{\noindent}
\begin{document}

\title{Mathematical Caricature of  Large Waves.}


\author{ Mikhail Kovalyov}{
  address={University College, Sungkyunkwan University,  300 Cheoncheon-dong, Jangan-gu, Suwon city, Gyeonggi-do, South Korea, 440-746.  E-mail: mkovalyo@gmail.com}
}

\begin{abstract}
The Kadomtsev-Petviiashvili equation is considered as a mathematical caricature of large and rogue waves.
\end{abstract}

\maketitle


  Large waves, also known as rogue, freak or monster waves, have been  in marine folklore since the time immemorial  inspiring  awe and fear.  Their description can be found  in Greek,  Scandinavian and Russian mythology,  in 1887-1898   E. Rudolph, a  professor of Geophysics of the University of Strasburg, Germany published   over 600 pages on large waves and seaquakes   documenting more than 550 drastic encounters of people with the sea. The stories of numerous sailors and sea men have  inspired many an artist, e.g.   Russian-Armenian  artist Ivan Aivazovsky dedicated his famous painting  "Devyatiy Val" (translates from Russian as "The ninth wall of water") to such waves,  Le Gray's 1857 photograph "The Great Wave" sold in 1999 for \$838,000.      Yet despite the overwhelming evidence the scientific establishment  pretty much  ignored large waves, any evidence presented was dismissed as un-scientific.  Relatively recently scientific establishment admitted that monster waves exist and a deluge of research papers followed. The search for their explanation  became so popular that    any  possible blow-up in a solution of an equation remotely related to waves is now called "a rogue" or "freak" wave and is considered to be an "explanation" regardless of whether it has anything to do with such waves or not.

\setlength\unitlength{1mm}
\vskip61mm
 \vbox{
\begin{picture}(0,0)(0,0)
\put(-2.1,2.5){\includegraphics[scale=.07]{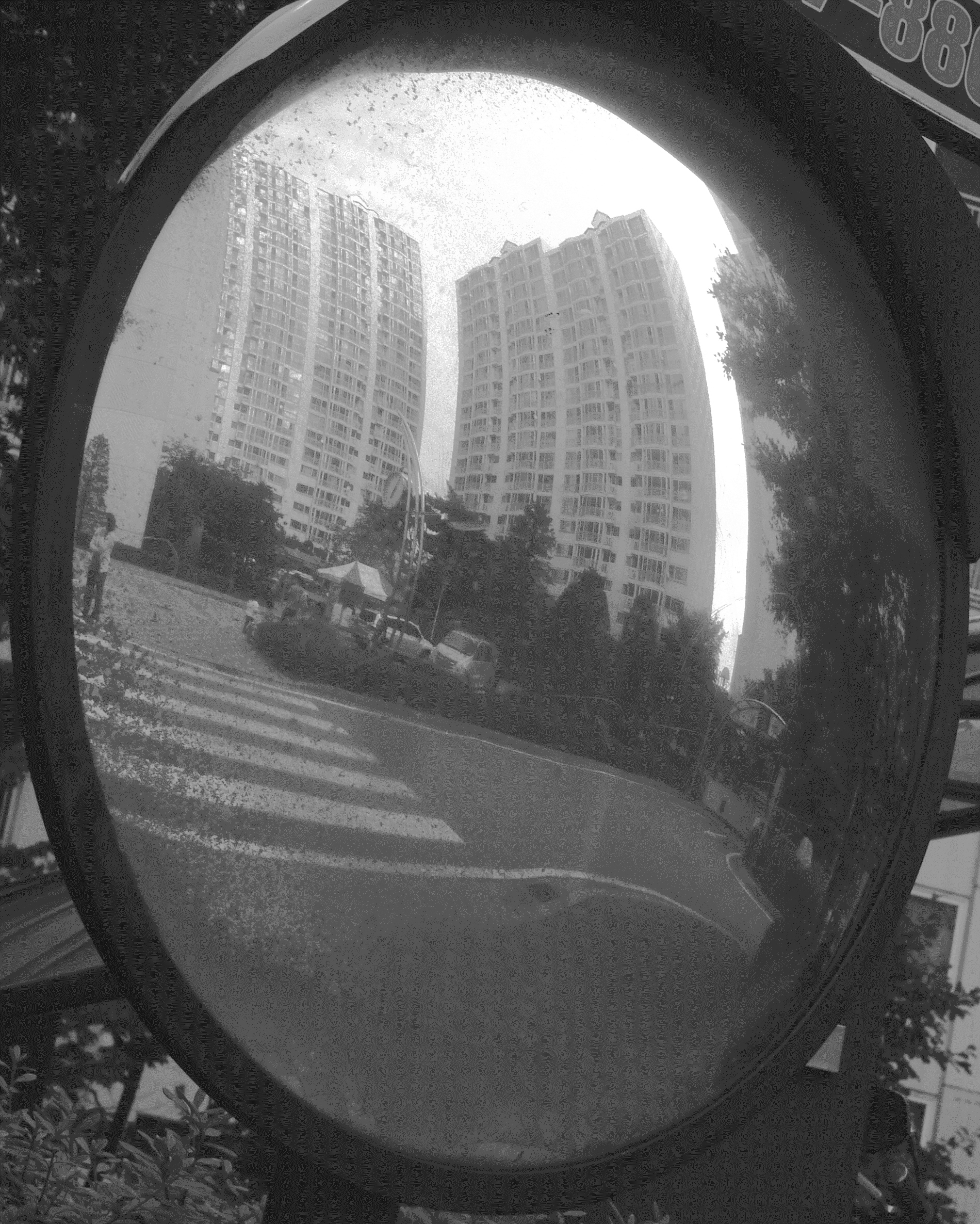}}
\end{picture}}

 \vskip-65mm \def\hang15{\hangindent15\parindent}
 \hang15{ \hskip55mm  In this talk we will look at large   waves through the mirror of the Kadomtsev-Petviashvili equation.  As pretty much any mathematical model  the Kadomtsev-Petviashvili equation may be compared to a curved mirror. A small fly flying by the mirror will see an almost perfect reflection of itself but the reflections of much larger buildings will be quite distorted. Yet despite the distortion the latter will show  windows, floors, etc. of the buildings. Similarly  the Kadomtsev-Petviashvili equation provides a good description   of soliton walls but   only a distorted description of larger waves;  however, despite the distortion  the appropriately chosen solutions of the Kadomtsev-Petviashvili equation show us some features of large waves.
 It would  be nice to have a better mathematical model  but as Albert Einstein once said, "As far as the laws of mathematics refer to reality, they are not certain; and as far as they are certain, they do not refer to reality ".
 
 \hskip55mm To derive  the Kadomtsev-Petviashvili equation in physical coordinates
$  x^\prime,y^\prime, z^\prime, t^\prime  $  we assume that the motion of fluid is restricted to the region   $-h< z^\prime < \epsilon h \eta(x^\prime,y^\prime,t^\prime)$, where
  $z^\prime=\epsilon h \eta(x^\prime,y^\prime,t^\prime)$ is the  elevation of the free surface, $h>0, \; 0<\epsilon \ll1  $ are constants. The
  velocity of fluid $\mathbf{u^\prime}$ has a potential }

\noi $\epsilon \phi$, i.e.    $ \mathbf{u^\prime}=\epsilon\nabla \phi.  $
In  dimensionless variables
$   T= t^\prime\sqrt{\frac{\epsilon^{3 }g }{h }},\;  X=\sqrt{\epsilon}\frac{x^\prime- t^\prime \sqrt{g  h}}{h } ,\;  Y= {\epsilon} \frac{y^\prime}{h },\;  Z=\frac{z^\prime}{h},\;   \Phi= \phi\sqrt{\frac{\epsilon }{gh^3}},\;  S=\frac{s}{\rho h^2 g},
 $ the equations of motion take form

$$  \hskip-40mm \begin{aligned}
&  \epsilon  \frac{\partial^2 \Phi}{\partial X^2}+   \epsilon^2   \frac{\partial^2 \Phi}{\partial Y^2}+\frac{\partial^2 \Phi }{\partial Z^2}=0 \hskip35mm \text{ in   } -1<Z<  \epsilon \eta ,\\
& \frac{\partial \Phi}{\partial Z}= 0\hskip72mm \mbox{ on }  Z = -1,
\\
& \dfrac{\partial \Phi}{\partial Z}  =-   \epsilon\dfrac{\partial \eta}{\partial X} + \epsilon^2    \dfrac{\partial \eta}{\partial T} + \epsilon^2 \dfrac{\partial \Phi}{\partial X} \dfrac{\partial \eta}{\partial X} + \epsilon^3 \dfrac{\partial \Phi}{\partial Y }\dfrac{\partial \eta}{\partial Y} \hskip13mm \mbox{ on  } Z = \epsilon  \eta ,\\
\end{aligned}$$

$$   \begin{aligned}&   \eta -S \epsilon\eta_{XX}-S\epsilon^2 \eta_{YY}  -  \tfrac{\partial \Phi}{\partial X}+\epsilon  \tfrac{\partial \Phi}{\partial T}+\tfrac 12    \left(\tfrac{\partial \Phi}{\partial Z}\right)^2
 +\tfrac 12 \epsilon   \left(\tfrac{\partial \Phi}{\partial X}\right)^2+  \tfrac 12 \epsilon^2  \left(\tfrac{\partial \Phi}{\partial Y}\right)^2=0   \mbox{ on }  Z = \epsilon  \eta,
\end{aligned}$$
with details of the derivation elaborated in [Kov 1].

Solutions are sought   in the form
$
\Phi=\Phi_0+\epsilon\Phi_1+\epsilon ^2 \Phi_2+\dots,
\newline \;\;\eta=\eta_0+\epsilon\eta_1+ \dots,
$
which upon substitution in the equations of motion yield
$$
\dfrac{\partial}{\partial x^\prime}\left[ \dfrac{1}{  \sqrt{  g h} } \;\eta_{0t^\prime}+  \eta_{0x^\prime}+ \dfrac{\rho g h^2-3S}{6\rho g  }\eta_{0x^\prime x^\prime x^\prime}+\dfrac {3\epsilon }{2}\eta_{0 }\eta_{0x^\prime}\right]+\dfrac{1 }{2 }\eta_{0y^\prime y^\prime} =0,
$$
$$
  u^\prime_{x^\prime} \hskip-1mm= \hskip-1mm\epsilon\sqrt{gh}\;\eta_0+o(\epsilon ),  \;\;
u^\prime_{z^\prime} \hskip-1mm=  \hskip-1mm -  \epsilon (h+z^\prime)\sqrt{gh}  \;\dfrac{\partial \eta_0}{\partial x^\prime}+o\left(\epsilon^{3/2}  \right).
$$
 Change of variables
$  \alpha^2=\dfrac{2\rho g  }{\rho g h^2-3S},    \; t^\prime=  \dfrac{3\alpha^2\; t }{\sqrt{gh}} ,   x^\prime=   x +{3\alpha^2} t ,\;y^\prime=\dfrac {y }{\sqrt{2}\thinspace} , \; \eta_0=\dfrac {4f}{3\epsilon\alpha^2}  ,
$
simplifies the  latter equations   to
$$
\dfrac{\partial}{\partial x}\Big[ f_t+f_{xxx}+6ff_x\Big]+3\alpha^2 f_{yy}=0,   \quad \alpha^2=\pm 1,
\eqno{\rm
(KP)}
$$
  known as {\it the Kadomtsev-Petviashvili   equation} or simply KP and
$$
  u^\prime_{x^\prime} \hskip-1mm= \hskip-1mm\;\dfrac {4\sqrt{gh} }{3 \alpha^2} f+o(\epsilon ),  \;\;
u^\prime_{z^\prime} \hskip-1mm=  \hskip-1mm -   (h+z^\prime) \;\dfrac {4\sqrt{gh} }{3 \alpha^2}\dfrac{\partial f }{\partial x^\prime}+o\left(\epsilon^{3/2}  \right),\eqno{\rm
(VC)}
$$
which we call the velocity conditions. The latter are often ignored yet, as will see later, they play fundamental role in the explanation of rogue waves.

The simplest   solutions of KP are the so-called multi-soliton solutions
$$
f(x,y,t)=2\frac{\partial^2}{\partial x^2}\ln \det{\pmb  A},
$$
where
$ \displaystyle
{\pmb  A}_{mn}=\delta_{mn}
+\frac{c_n  \enskip e^{(p_n+q_n)x+(q_n^2-p_n^2)  y-(p_n+q_n)\big[(p_n +q_n)^2+3\alpha^2 (p_n-q_n)^2\big]t}}{p_n+q_m},
$
$m,n=1\dots N,   \delta_{mn}  $ are the regular Kronecker symbols, $c_n,\; p_n,\; q_n   $ are     real constants.  Numerous illustrations of soliton  and multi-soliton waves are available on the Internet.

  If we take nonlinear superposition of 2N solitons with $
p_{2k-1}={\overline {p_{2k}}}=i \lambda_k+\mu_k +\varepsilon e^{i \chi_n}, \;
q_{2k-1}={\overline{ q_{2k}}}=i \lambda_k-\mu_k +\varepsilon e^{-i \chi_n}, \;
c_{2k-1}={\overline{ c_{2k}}}=2\varepsilon e^{i\gamma_k-i\chi_n+\varrho_k\varepsilon},\;
\lambda_k,\mu_k,\gamma_k,\varrho_k\in{  R},$ $\; k=1\dots N
$
 and let $\epsilon \to 0$ we obtain

$$
u(x,y,t)=2\frac{\partial^2}{\partial x^2}\ln  \det{\pmb  K},\eqno{\rm
(1)}
$$
where
$
\;\; {\pmb  K}=\left(
\begin{array}{ccccc}
K_{11}&K_{12} &\dots& K_{1N} \\
K_{21} &K_{22}&\dots& K_{2N} \\
\vdots&\vdots&\vdots&\vdots \\
K_{N1} &K_{N2} &\dots &K_{NN}
\end{array}\right); \hskip5mm
 K_{nn}= \Upsilon_n  - \dfrac{\sin2\Gamma_n}{2\lambda_n}; \hskip5mm  \text { for }  n\ne k \hskip4mm K_{nk}=
 \left[
 \dfrac{(\lambda_n-\lambda_k)\sin(\Gamma_n-\Gamma_k)}{\alpha^2(\mu_n-\mu_k)^2+
(\lambda_n-\lambda_k)^2}-  \right. \newline \left. \dfrac{(\lambda_n+\lambda_k)\sin(\Gamma_n+\Gamma_k)}{\alpha^2 (\mu_n-\mu_k)^2+(\lambda_n+\lambda_k)^2}\right]
+    \alpha     \left[  \dfrac{(\mu_n-\mu_k)\cos(\Gamma_n+\Gamma_k)}{\alpha^2(\mu_n-\mu_k)^2+(\lambda_n+\lambda_k)^2}
   -
\dfrac{(\mu_n-\mu_k)\cos(\Gamma_n-\Gamma_k)}{\alpha^2 (\mu_n-\mu_k)^2+(\lambda_n-\lambda_k)^2}\right];  \hskip5mm
\Gamma_n=  \gamma_n+\lambda_n x-2\lambda_n\mu_n y+4\lambda_n(\lambda_n^2-3\alpha^2\mu_n^2)t; \hskip5mm  \Upsilon_n=\rho_n+x\cos (\alpha\chi_n) +2\left[ \dfrac{\lambda_n\sin (\alpha\chi_n)}{\alpha}-\mu_n\cos (\alpha\chi_n) \right]y + 12\Big[\lambda_n^2\cos(\alpha\chi_n)-\alpha^2\mu_n^2\cos (\alpha\chi_n) +2\alpha \lambda_n\mu_n\sin (\alpha\chi_n)\Big]t;
\hskip5mm \lambda_n \textmd{{'s, }} \mu_n\textmd{'s, } \chi_n\textmd{'s, } \gamma_n\textmd{'s, } \rho_n\textmd{'s }$ are   constants.

For $N=1$ formulas (1) degenerate into
$$
f(x,y,t)=2\displaystyle\frac{\partial^2}{\partial x^2}\ln  \Big[2\lambda\Upsilon   - {\sin 2\Gamma } \Big],
\eqno{\rm
(2)}
$$
where $\displaystyle \Gamma=  \gamma+\lambda x-2\lambda\mu y+4\lambda(\lambda^2-3\alpha^2\mu^2)t;\hskip5mm
  \Upsilon =\rho +x\cos (\alpha\chi) +2\Bigg[ \dfrac{\lambda\sin (\alpha\chi)}{\alpha}-\mu\cos (\alpha\chi) \Bigg]y +  12\Big[\lambda^2\cos(\alpha\chi)-\alpha^2\mu^2\cos (\alpha\chi) +2\alpha \lambda\mu\sin (\alpha\chi)\Big]t;\hskip5mm \lambda, \mu, \chi, \gamma, \rho  $ are   constants.  We shall refer to (2) as a {\it harmonic breather} for reasons discussed in [Kov 2], formula (1) may be viewed as nonlinear superposition of harmonic breathers.

Solution (2) of KP  is  singular and   approaches $-\infty $  as we move towards the singular curve $2\lambda\Upsilon -  {\sin  2\Gamma }=0.  $  To be physically plausible it needs to be regularized to remove infinities; since there is no clear way to determine what kind of regularization is most suitable we choose   an ad-hoc regularization
$$
f \to F= \displaystyle\frac{f}{\vert f\vert }\ln \Big[\ln \Big(\Big \vert e^f-1 \Big\vert+1\Big)+1\Big] \eqno{\rm
(3)}
$$
due to its properties
$
  F  \approx f, \textmd{ if }  0< \vert f\vert \ll 1;\;\;\;
 F=0, \textmd{ if } f=0; \;\;\;
F\to -\ln (\ln 2+1),  \textmd{ as } f \to -\infty; \;\; \;F \approx  \ln f \textmd{ as } f \to +\infty.
 $
The regularization is close to the original function  for $\vert f\vert \ll 1,$     removes   negative infinity  and  reduces  large amplitudes to make them more

\noi physically plausible.

Regularization  of harmonic breather (2) with $\chi=0 $ is shown in Figure 1. Harmonic breather (2) with $\chi=0 $ is divided into two simply connected components by a  singular line suggesting that each component might describe a separate physical wave, regularization of one such half is shown in Figure 2. The physical waves described, albeit imperfectly,  by such simply connected components  are known as undular bores, a few examples are  shown in Figures 3-6;  more pictures of undular bores can be found on the Internet.

\begin{figure}[!h]
  \centerline{\includegraphics[scale=.25]{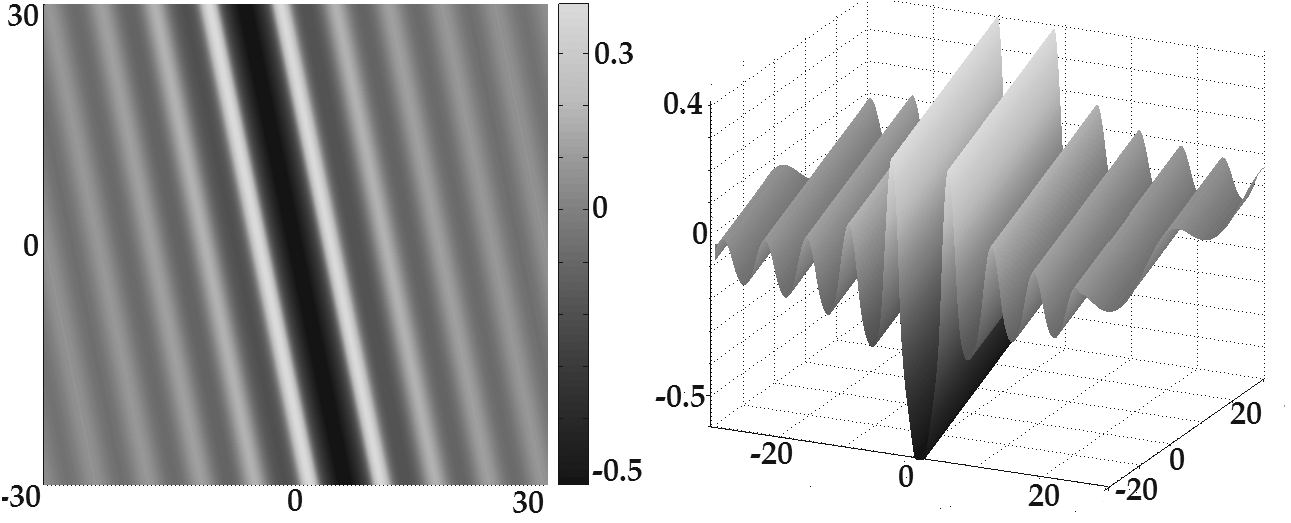}}
\caption{\hskip-2mm Graph of  regularization (3) of (2) with $\alpha\hskip-0.5mm=\hskip-0.5mm1,  \;  \lambda\hskip-0.5mm=\hskip-0.5mm0.5,\;\mu\hskip-0.5mm=\hskip-0.5mm-0.1,\;
\gamma\hskip-0.5mm=\hskip-0.5mm0,\rho\hskip-0.5mm=\hskip-0.5mm0,\; \chi\hskip-0.5mm=\hskip-0.5mm0,\; t\hskip-0.5mm=\hskip-0.5mm0.  $ }
\end{figure}
\begin{figure}[!h]
 \centerline{\includegraphics[scale=.25]{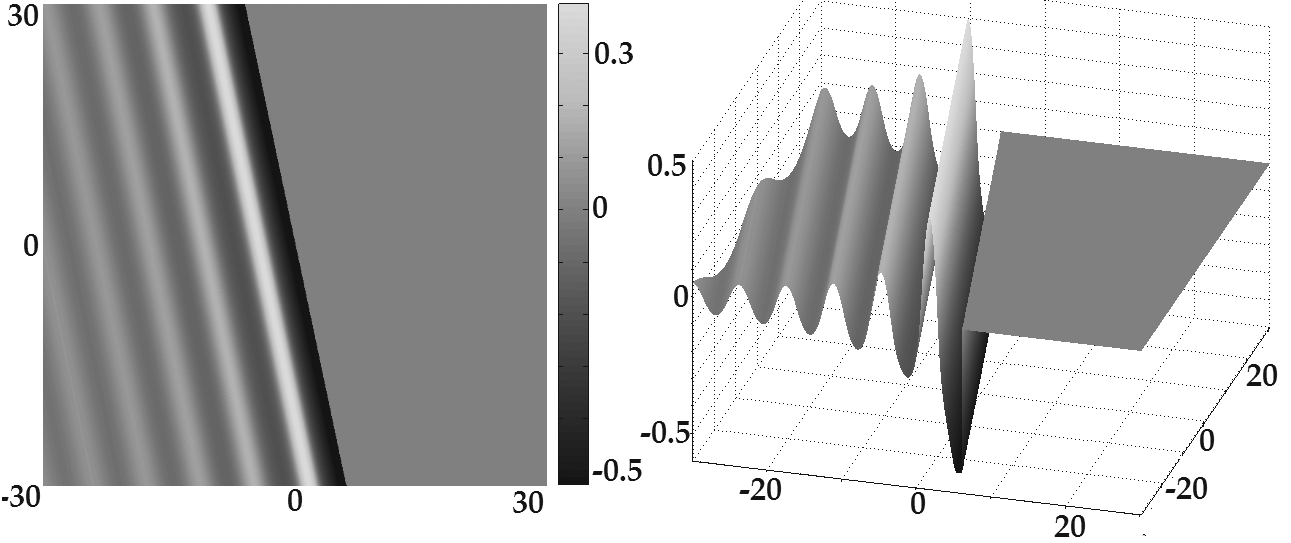}}
\caption{\hskip-2mm Graph of  regularization (3) of a simply connected half of  (2) with $\alpha\hskip-0.5mm=\hskip-0.5mm1,  \;  \lambda\hskip-0.5mm=\hskip-0.5mm0.5,\;\mu\hskip-0.5mm=\hskip-0.5mm-0.1,\;
\gamma\hskip-0.5mm=\hskip-0.5mm0,\rho\hskip-0.5mm=\hskip-0.5mm0,\; \chi\hskip-0.5mm=\hskip-0.5mm0,\; t\hskip-0.5mm=\hskip-0.5mm0.  $ The plot is discontinuous  along the straight line   $2\lambda\Upsilon   - {\sin 2\Gamma }=0 $, however Matlab replaced the discontinuity   with a vertical strip and the author decided to leave it that way. }
\end{figure}

\vskip-1mm\begin{figure}[!h]
  \put(0,-0.9)\centerline{\includegraphics[scale=.23]{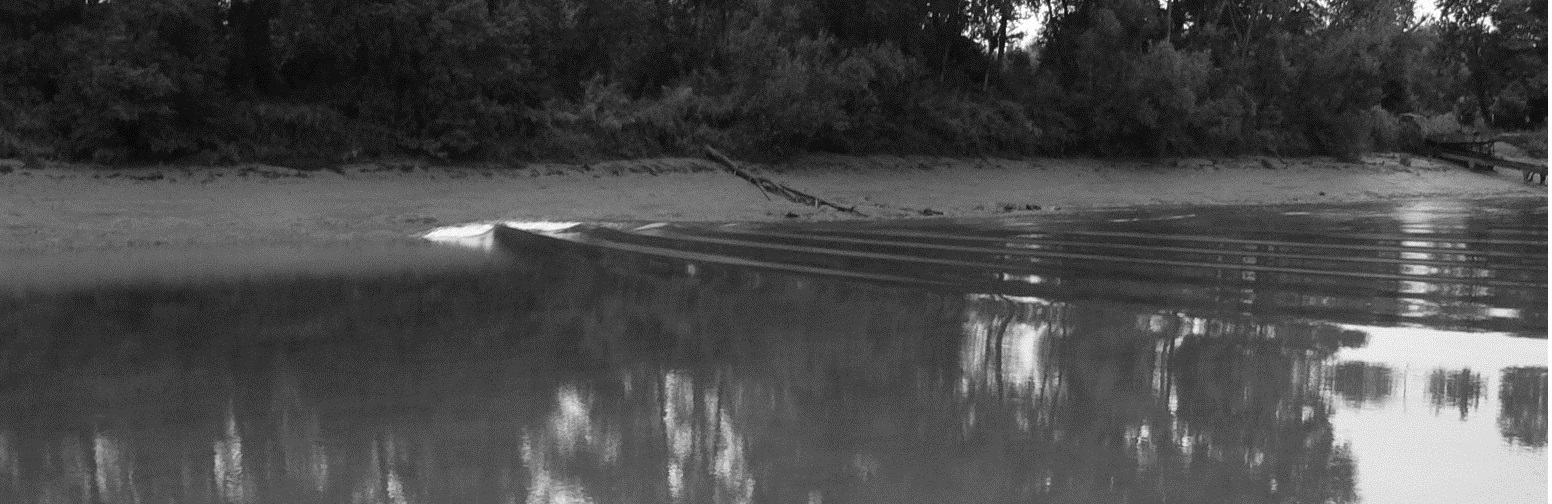}\caption{An undular bore generated by tide on the Garonne river  at Arcins on  July 6, 2008 around 07:10 views from right bank.    Courtesy of Hubert Chanson.}}
\end{figure}

\begin{figure}[ht]
  \centerline{\includegraphics[scale=.38]{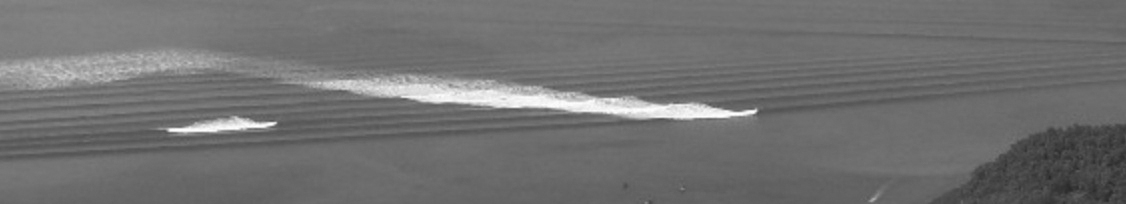}\caption{An undular bore generated by the tsunami of December 26, 2004 approaching Koh Jum island, off the coast of Thailand, after it passed around the Phi
Phi islands.      Copyright Anders Grawin, 2006. Reproduced from http://www.kohjumonline.com/anders.html with permission.}}
\end{figure}

\hskip-3mm
\begin{figure}[!h]
  \centerline{\includegraphics[scale=.66]{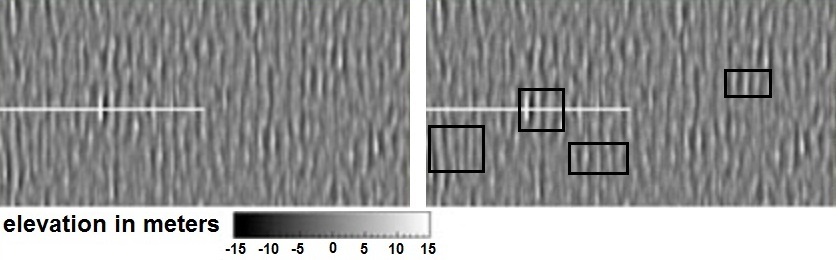}\caption{ Image captured by ESA, courtesy of ESA and Deutsches Zentrum fur Luft- und
Raumfahrt (DLR). On the left is the original image, on the right is the same image with some
waves of the type shown in Figures 1 and 2 enclosed in black rectangles. The middle of the
horizontal white line marks the wave with the highest crest, it is the leading crest of a wave of
the type shown in Figure 2. Reproduced with permission.}}
\end{figure}~\begin{figure}[!h]
  \centerline{\includegraphics[scale=.35]{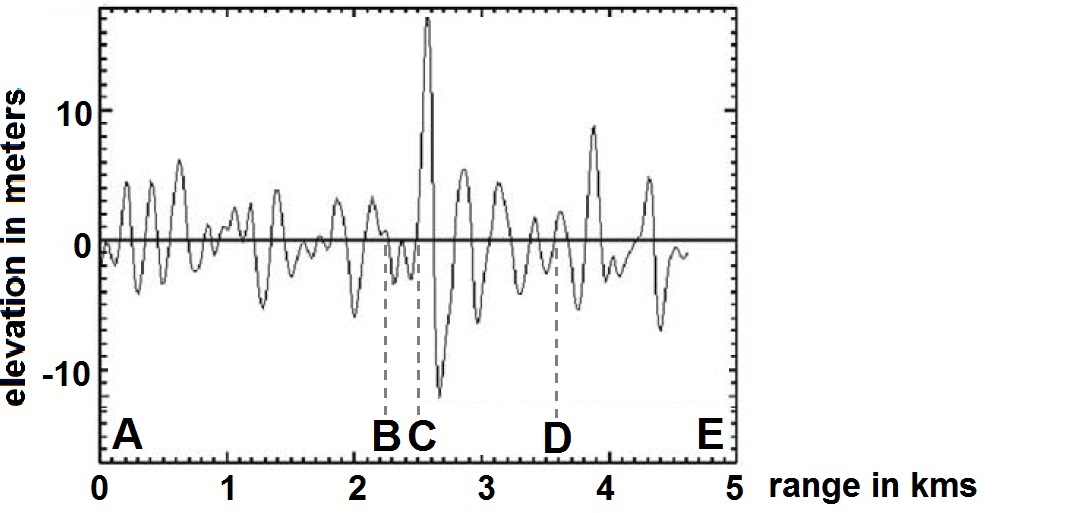}\caption{ The vertical transect of the previous picture  along the straight white line.  Region CD is as predicted by the formula,  region BC corresponds to the negative singularity. In regions AB and DE the undular bore-like behavior is masked  by other waves. Reproduced with permission.}}
\end{figure}

Typical undular bore exhibits non-smooth behavior at the leading  front reflected by the presence of a singularity in the solution. Formulas  (VC)   imply wave overturning, large acceleration and thus the presence of large forces at the leading front/trough; description of the power of undular bores near the leading front/trough is given in  Wolanski, E., Williams, D., Spagnol, S., and Chanson, H. (2004),
 Undular
Tidal Bore Dynamics in the Daly Estuary, Northern Australia. Estuarine, Coastal and Shelf Science, Vol. 60, No. 4, pp. 629-636 (DOI: 10.1016/j.ecss.2004.03.001)  , "{ This unsteady motion was sufficiently energetic to topple moorings that had survived much higher, quasi-steady currents of 1.8 m/s}"

Regularization  of harmonic breather (2) with $\chi\neq 0 $ is shown in Figure 7 while physical waves described, albeit imperfectly, by such waves are shown in Figures 8-9.

\begin{figure}[!h]
  \centerline{\includegraphics[scale=.25]{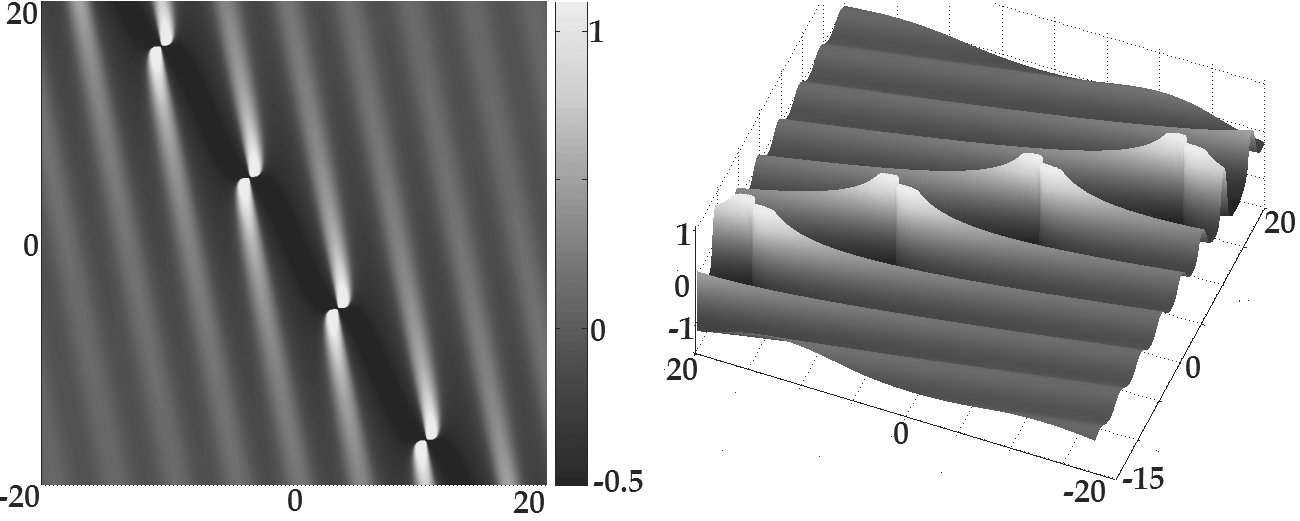}\caption{\hskip-3mm Graph of regularization  (3) of (2) with $\alpha\hskip-0.55mm=\hskip-0.55mm1,     \lambda\hskip-0.55mm=\hskip-0.55mm0.65, \mu\hskip-0.55mm=\hskip-0.55mm-0.1,
  \gamma\hskip-0.55mm=\hskip-0.55mm0, \rho\hskip-0.55mm=\hskip-0.55mm0,
 \chi\hskip-0.55mm=\hskip-0.55mm.105\pi,  t\hskip-0.55mm=\hskip-0.55mm0. $}}
\end{figure}

    \begin{figure}[!h]
 {\includegraphics[scale=.6]{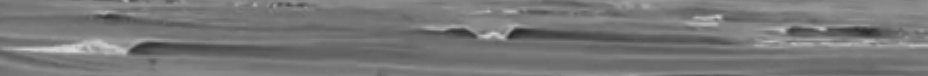}\caption{Portion of time frame t = 2 : 11 from the video at [Rip Curl] depicts a wave on the river of Seven Ghosts, Indonesia. The shape of
the wave is in agreement with Figure 2 with $\chi$ close to but not equal to zero except for wave
overturning exhibited by the physical wave shown but not by Figure 7. Reproduced with permission. }}
\end{figure}

  \begin{figure}[!h]
 {\includegraphics[scale=.25]{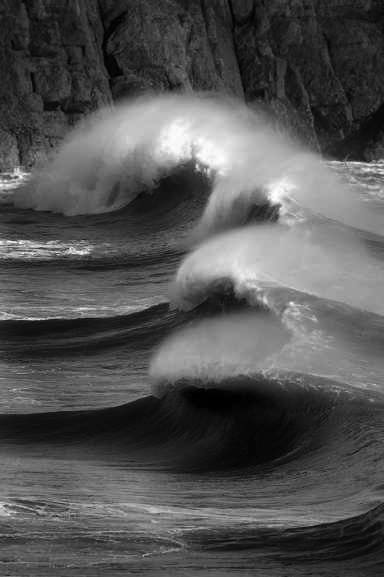}\caption{ The shape of this  wave at Hells Mouth, Wales,    is in agreement with Figure 2 with $\chi\neq 0$  except for wave
overturning exhibited by the physical wave shown but not by Figure 7. The photograph is reproduced from  http://magicseaweed.com/photoLab/viewPhoto.php?photoId=18664 , courtesy of John Wormald. }}
\end{figure}

Harmonic
breather (2) with $ \chi \neq 0$    is divided into two simply connected by the singular curve  suggesting that each  simply connected component  might describe a separate physical  wave, regularization of one such component
is shown in Figure 10 while  a corresponding physical wave is shown in Figure~11.

   \begin{figure}[h!]
 {\includegraphics[scale=.32]{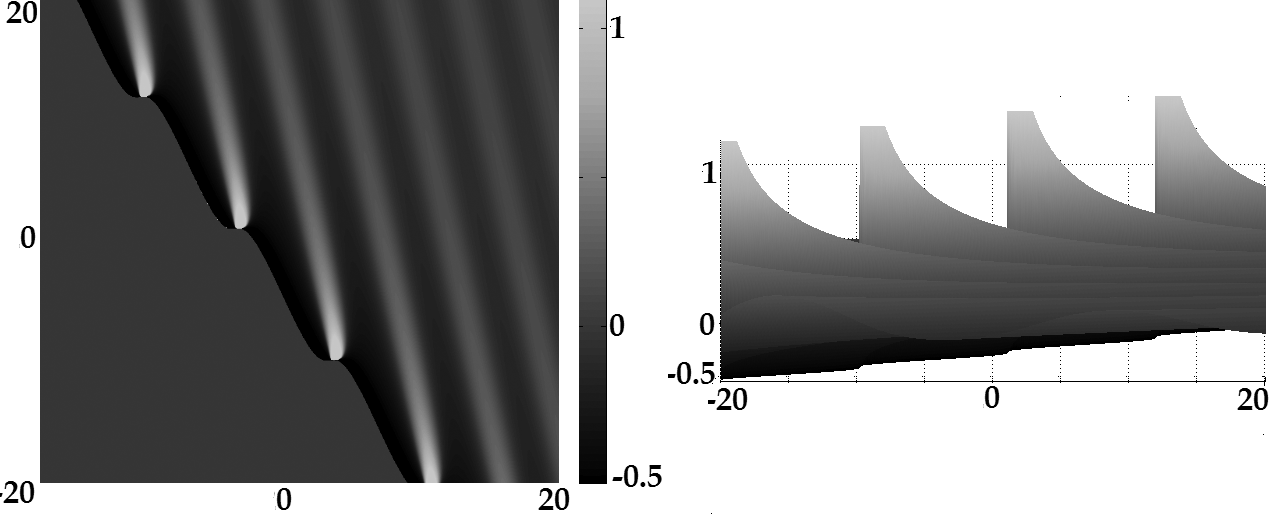}\caption{Graph of regularization  (3) of  a simply connected component of (2)  with  $\alpha\hskip-0.5mm=\hskip-0.5mm1,  \;  \lambda\hskip-0.5mm=\hskip-0.5mm0.65,\;\mu\hskip-0.5mm=\hskip-0.5mm-0.1,\;
\gamma\hskip-0.5mm=\hskip-0.5mm0,\rho\hskip-0.5mm=\hskip-0.5mm0,\; \chi\hskip-0.5mm=\hskip-0.5mm0.105\pi ,\; t\hskip-0.5mm=\hskip-0.5mm 0.7.$ }}
\end{figure}

   \begin{figure}[!h]
 {\includegraphics[scale=1.0]{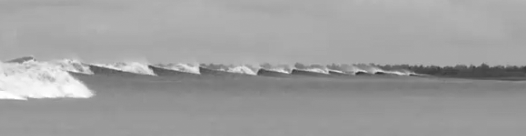}\caption{Portion of time frame t =1:17 from the video at [Rip Curl] depicts a wave on the river
of Seven Ghosts, Indonesia, its shape  is in agreement with Figure 10 with $\chi\neq0$  except for wave
overturning exhibited by the physical wave shown but not by Figure 10. Reproduced with permission.}}
\end{figure}

  If   the spectral parameters $\lambda, \mu$ are sufficiently large the graph of the regularization of a  harmonic breather with $\chi\neq 0$ is shown in Figure 12 while a corresponding physical wave is shown in Figure 13.

   \begin{figure}[!h]
 {\includegraphics[scale=.4]{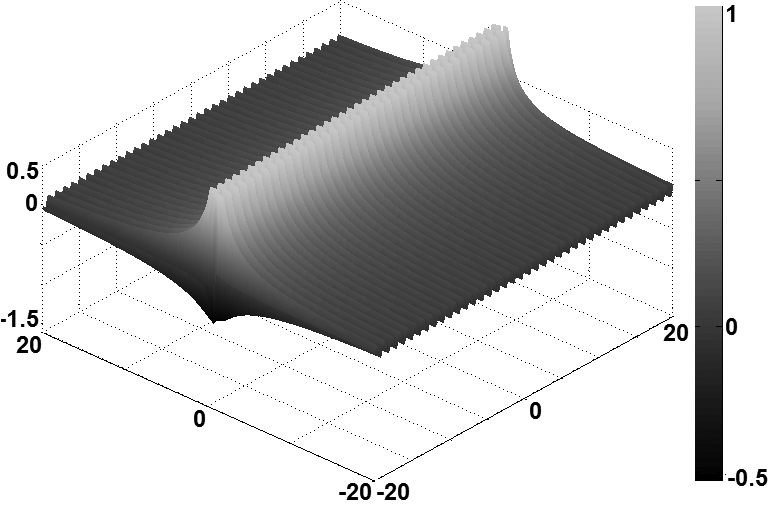}\caption{Graph of regularization  (3) of    (2)  with  $\alpha\hskip-0.5mm=\hskip-0.5mm1,  \;  \lambda\hskip-0.5mm=\hskip-0.5mm3,\;\mu\hskip-0.5mm=\hskip-0.5mm0,\;
\gamma\hskip-0.5mm=\hskip-0.5mm0,\rho\hskip-0.5mm=\hskip-0.5mm0,\; \chi\hskip-0.5mm=\hskip-0.5mm0.55\pi ,\; t\hskip-0.5mm=\hskip-0.5mm 0.$ }}
\end{figure}

  \begin{figure}[!h]
 {\includegraphics[scale=.5]{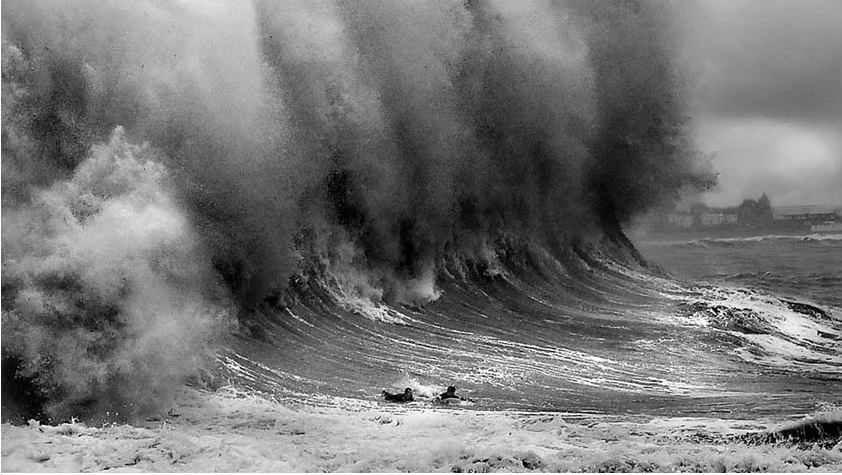}\caption{A physical wave with the shape predicted by Figure 12. Notice the streaks across the wave and a "cloud" of droplets. Compare it to the wave shown in Figure 16. Reproduced from public domain http://mpora.com/photos/T5SsfsQ7BR . }}
\end{figure}

  Another class of explicit solutions of KP  is obtained by  substitution
 $
   \lambda_k \to i\lambda_k, \; \chi_k \to i\chi_k    , \; \gamma_k  \to i\gamma_k
 $ into formulas (1) and (2) yielding  correspondingly
 $$
u(x,y,t)=2\frac{\partial^2}{\partial x^2}\ln  \det{\pmb  K}
\eqno{{\rm(4a)}}
$$
where   ${\pmb  K}=\left(
\begin{array}{llll}
K_{11}&K_{12} &\dots& K_{1N} \\
K_{21} &K_{22}&\dots& K_{2N} \\
\vdots&\vdots&\vdots&\vdots \\
K_{N1} &K_{N2} &\dots &K_{NN}
\end{array}\right);\hskip5mm K_{nn}= \Upsilon_n  - \dfrac{\sinh 2\Gamma_n}{2\lambda_n}; \hskip5mm \text{ for } n \neq k \hskip3mm K_{nk}=   \left[-
 \dfrac{(\lambda_n-\lambda_k)\sinh(\Gamma_n-\Gamma_k)}{\alpha^2(\mu_n-\mu_k)^2-
(\lambda_n-\lambda_k)^2} \right. \newline \left. + \dfrac{(\lambda_n+\lambda_k)\sinh(\Gamma_n+\Gamma_k)}{\alpha^2 (\mu_n-\mu_k)^2-(\lambda_n+\lambda_k)^2}\right]
+    \alpha     \left[  \dfrac{(\mu_n-\mu_k)\cosh(\Gamma_n+\Gamma_k)}{\alpha^2(\mu_n-\mu_k)^2- (\lambda_n+\lambda_k)^2}
- \dfrac{(\mu_n-\mu_k)\cosh(\Gamma_n-\Gamma_k)}{\alpha^2 (\mu_n-\mu_k)^2-(\lambda_n-\lambda_k)^2}\right]; \hskip5mm \Gamma_n=  \gamma_n+\lambda_n x-2\lambda_n\mu_n y-4\lambda_n(\lambda_n^2+3\alpha^2\mu_n^2)t;\hskip5mm \Upsilon_n=\rho_n+x\cosh (\alpha\chi_n) -2\Big[\displaystyle\frac{\lambda_n\sinh (\alpha\chi_n)}{\alpha}+\mu_n\cosh (\alpha\chi_n) \Big]y -  12\Big[\lambda_n^2\cosh(\alpha\chi_n)+\alpha^2\mu_n^2\cosh (\alpha\chi_n) \newline + 2\alpha \lambda_n\mu_n\sinh (\alpha\chi_n)\Big]t;\hskip5mm \lambda_n\textmd{'s, } \mu_n\textmd{'s, } \chi_n\textmd{'s, } \gamma_n\textmd{'s, } \rho_n\textmd{'s  }$ are   constants.

 For $N=1$ formulas (4) degenerate to
 $$
 f(x,y,t)=2\displaystyle\frac{\partial^2}{\partial x^2}\ln  \Big[2\lambda\Upsilon   - {\sinh 2\Gamma } \Big],  \eqno{\rm(5)}
$$
where $ \Gamma=  \gamma+\lambda x-2\lambda\mu y-4\lambda(\lambda^2+3\alpha^2\mu^2)t;\hskip5mm \Upsilon =\rho +x\cosh (\alpha\chi) -2\Bigg[ \dfrac{\lambda\sinh (\alpha\chi)}{\alpha}+\mu\cosh (\alpha\chi) \Bigg]gy - 12\Big[\lambda^2\cosh(\alpha\chi)+\alpha^2\mu^2\cosh (\alpha\chi) +2\alpha \lambda\mu\sinh (\alpha\chi)\Big]t; \hskip3mm \lambda, \mu, \chi, \gamma,  \rho  $ are    constants.  We shall refer to (5) as {\it solitonic dipole} for reasons discussed in [Kov 2]. Regularization of a solitonic dipole  (5) with $\chi \neq 0$ is shown in Figure 14.     We can also take a half-wave as before although now we cut it in half in the middle not along the singular curve. Physical analogues of such waves are shown in Figures 16-18, more illustrations of physical waves described by solitonic dipoles may be found on the Internet.

   \begin{figure}[!h]
 {\includegraphics[scale=.25]{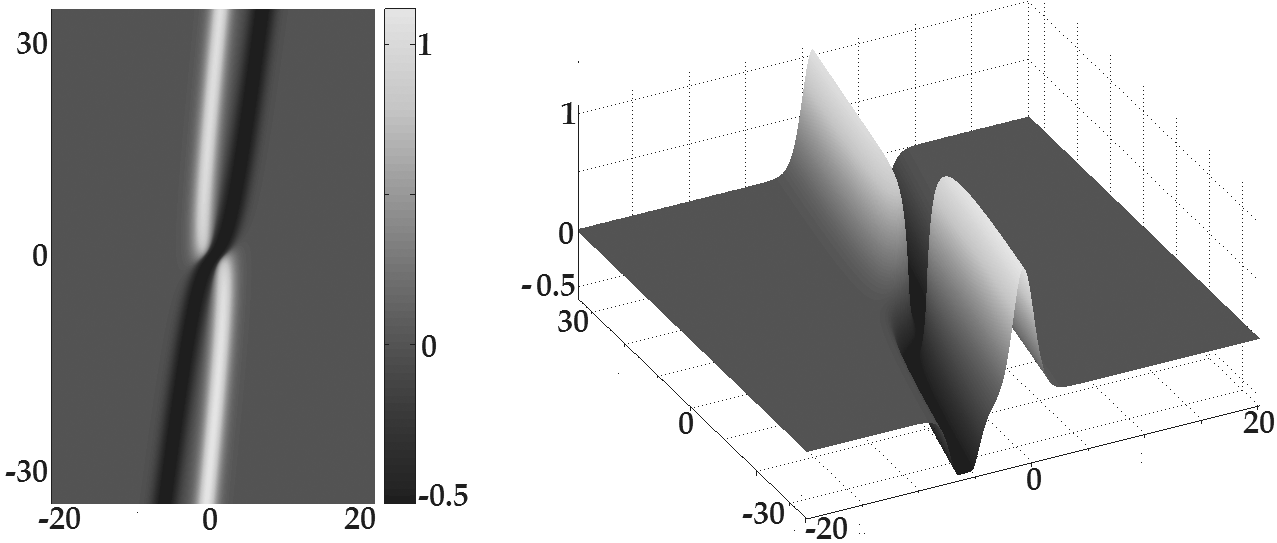}\caption{\hskip-2mm Graph of regularization  (3) of (5) with $\alpha\hskip-0.5mm=\hskip-0.5mm1,  \,  \lambda\hskip-0.5mm=\hskip-0.5mm1,\,\mu\hskip-0.5mm=\hskip-0.5mm0.05,\,
  \gamma\hskip-0.5mm=\hskip-0.5mm0,\,\rho\hskip-0.5mm=\hskip-0.5mm0,
\, \chi\hskip-0.5mm=\hskip-0.5mm0.6,\, t\hskip-0.5mm=\hskip-0.5mm0. $   }}
\end{figure}

    \begin{figure}[!h]
 {\includegraphics[scale=.25]{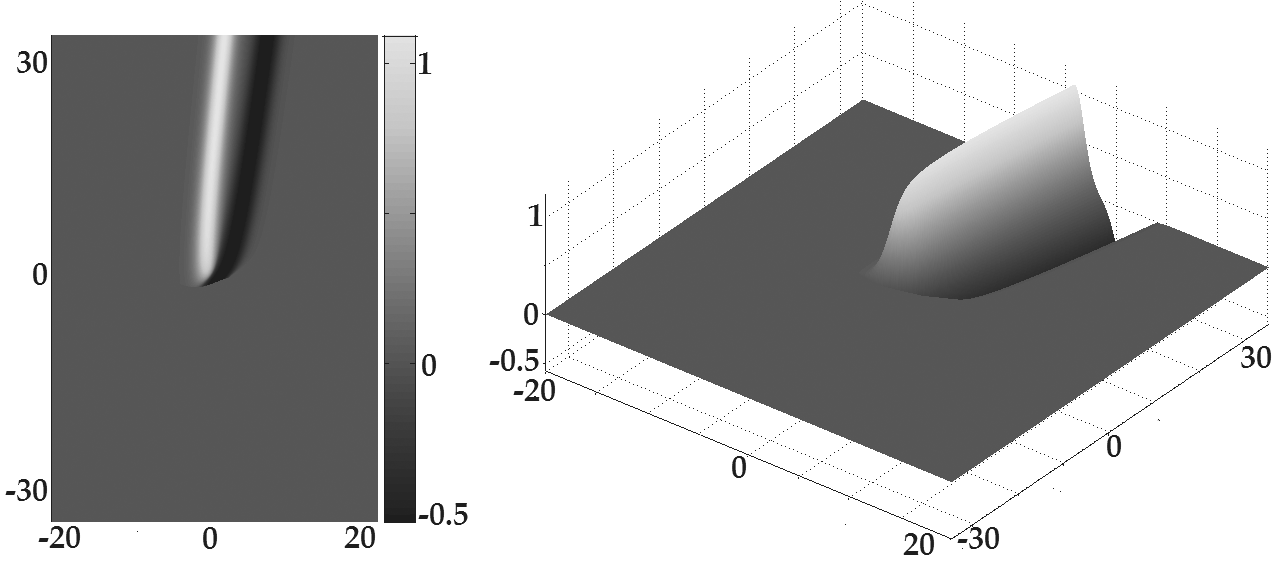}\caption{\hskip-2mm  Graph of regularization  (3) of the upper half of (5) with $\alpha\hskip-0.5mm=\hskip-0.5mm 1,  \,  \lambda\hskip-0.5mm=\hskip-0.5mm1,\,\mu\hskip-0.5mm=\hskip-0.5mm0.05,\,
\gamma\hskip-0.5mm=\hskip-0.5mm0,\,\rho\hskip-0.5mm=\hskip-0.5mm0,\,  \chi\hskip-0.5mm=\hskip-0.5mm0.6,\, t\hskip-0.5mm=\hskip-0.5mm0.
  \hskip110mm$  }}
\end{figure}

      \begin{figure}[!h]
 {\includegraphics[scale=.50]{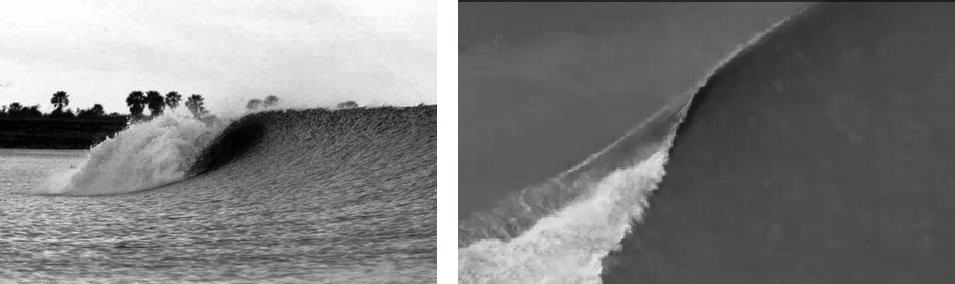}\caption{ On the left  is a tidal bore in the Amazon River; with the height reaching  up to 4 meters it  travels as much as 13 kilometers inland upstream from  the Atlantic Ocean in   February and March.  Reproduced from public domain http://captainkaisworld.blogspot.ca/2011/04/surfing-rivers-7-ghosts-severn-chinas.html .
  On the right is a portion  of time frame t =0:52 from the video at [Rip Curl] depicting a wave on the river
of Seven Ghosts, Indonesia. Unlike the wave shown in Figure 13 these waves do not have strongly defined transverse streaks or the "cloud" of droplets above it, the crest is relatively well-defined compared to Figure 13.   Reproduced with permission. }}
\end{figure}

      \begin{figure}[!h]
 {\includegraphics[scale=.22]{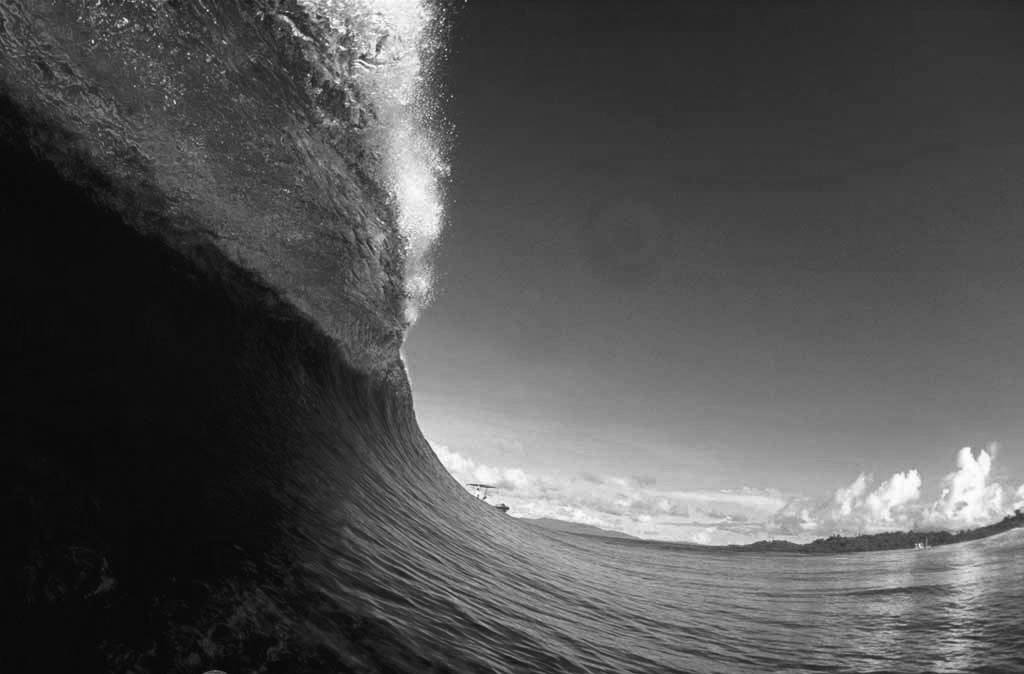}\caption{   The picture emphasizes the trough rather than the crest of the wave. Reproduced from a public web site, due to numerous web sites showing this picture the origin is unknown.   }}
\end{figure}

        \begin{figure}[!h]
 {\includegraphics[scale=.66]{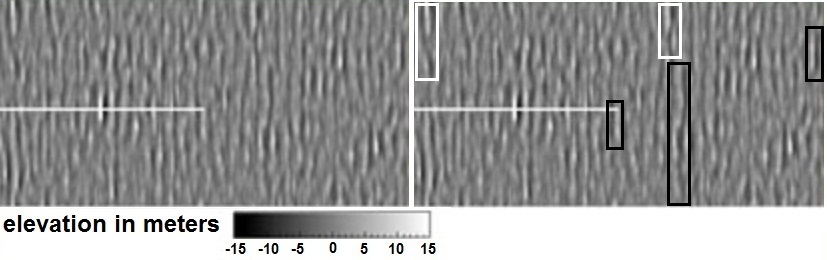}\caption{ Image  captured by ESA, courtesy of ESA and Deutsches Zentrum fur Luft- und Raumfahrt (DLR).  On the left is the original image, on the right is the same image with some half-waves   shown in   black rectangles and two full waves   shown  in white rectangles. Reproduced with permission. }}
\end{figure}

The waves discussed have relatively long life-span, yet there are accounts of extremely powerful waves that appear seemingly out of nowhere and disappear shortly thereafter into nowhere.  Remarkably,  KP predicts the existence of such waves and provides their explanation. Let us look at Figure 19 showing
$  \widetilde{f}(x,y,t)=\left\{\begin{array}{l} \displaystyle\frac{f}{\vert f\vert }\ln \Big[\ln \Big(\Big \vert e^f-1 \Big\vert+1\Big)+1\Big], \textmd{ if } f\leqslant 0,\\\\ \max\{ f, 10\},\textmd{ if } f>0,  \end{array} \right.,  $ where $f$ is nonlinear superposition of two solitonic dipoles given by (4); it illustrates
 the creation of a short-lived physically localized rogue wave  seemingly out of nowhere.  The physics behind it is shown in Figure 20.

        \begin{figure}[!h]
 {\includegraphics[scale=.36]{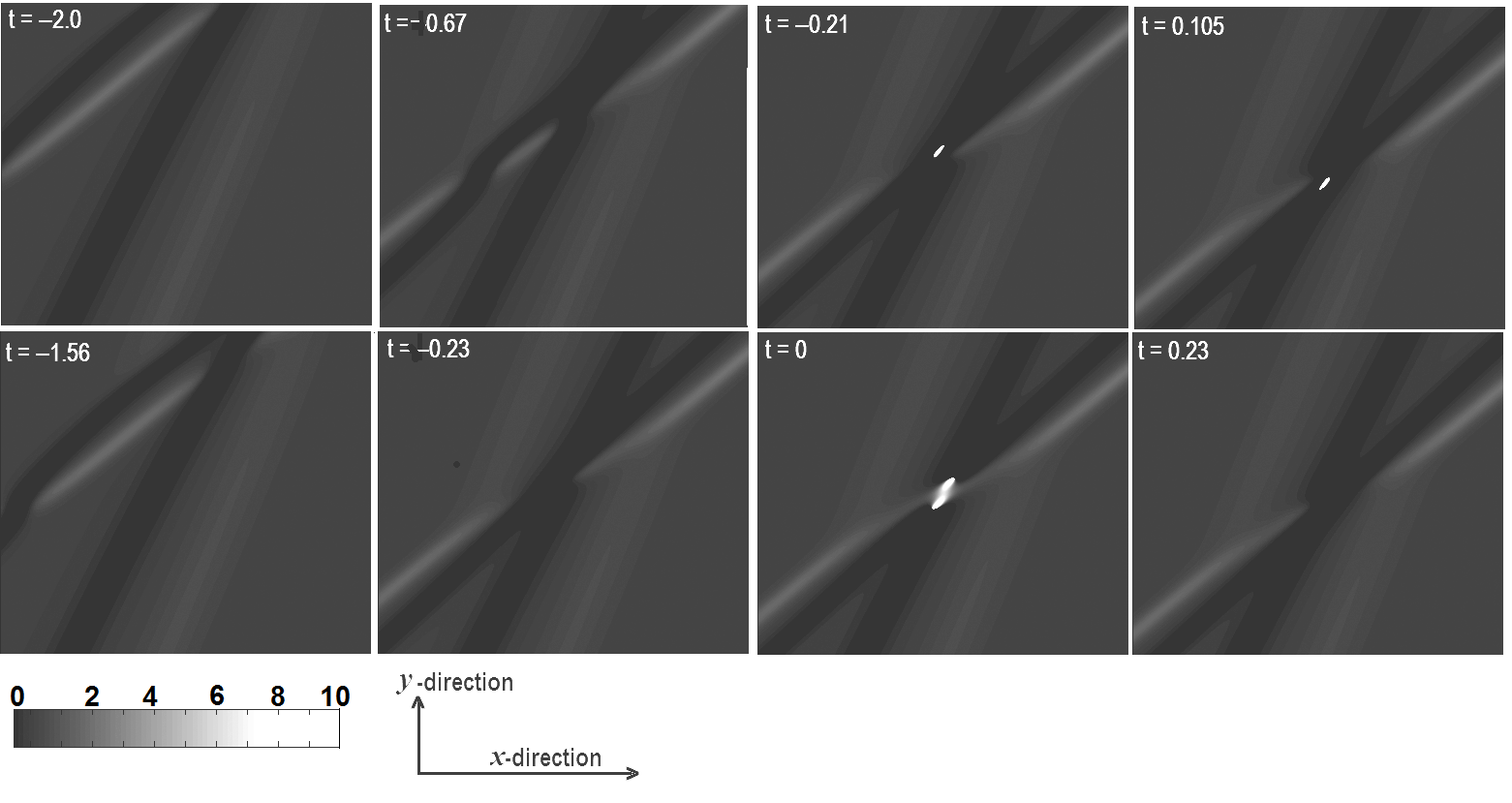}\caption{ Time evolution of $\widetilde{f}$   where $f$ is given by
(4) with
$N\hskip-1mm=\hskip-1mm2, \, \alpha\hskip-1mm=\hskip-1mm1,   \chi_1\hskip-1mm=  0.6,\,
\lambda_1\hskip-1mm= 0.5,  \,
\mu_1 \hskip-1mm= 0.2,\,
\gamma_1\hskip-1mm=  0,\,
\rho_1\hskip-1mm= 0,\,
\chi_2\hskip-1mm=\hskip-1mm - 0.7 ,\,
\lambda_2\hskip-1mm= 1  ,\,
\mu_2\hskip-1mm=0.5 ,\,
\gamma_2\hskip-1mm=  0,\,
\rho_2\hskip-1mm= 0   $  in the region
 $-15< x,y<15 $.   Notice the    rogue wave    shown by   white spots  near the center; it  is short-lived and exists only for t close to 0.  }}
\end{figure}
   \begin{figure}[!h]
 {{\includegraphics[scale=.47]{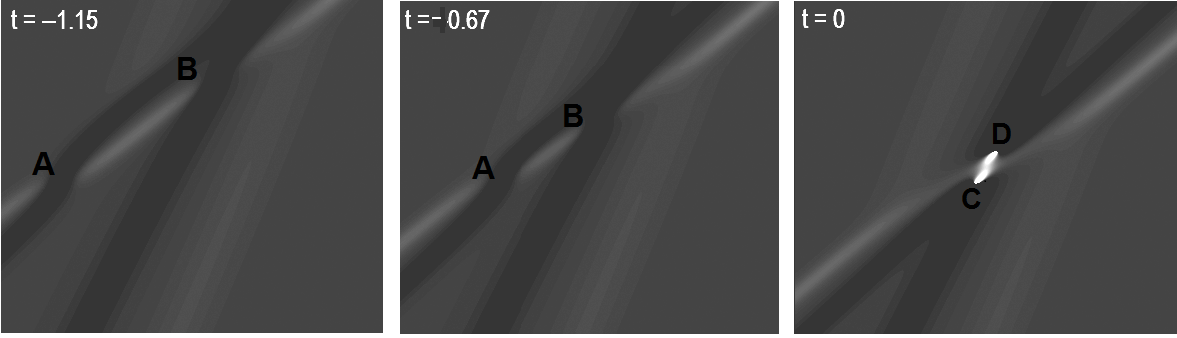}\caption{ Function of Figure 19 at $ t=-1.15, \; t=-0.67, \; t=0.$ According to (VC)  water in the white regions moves to the right while at the dark
regions it moves to the left thus creating undertow currents at points A and B. As   $t \to 0^-$  points A and B    move so close to each other that the flow of water from each current gets in the way of the flow in the other current, the two colliding flows have no other way to go but up creating a   rogue wave.  Once $t$ passes  $0$  points A and B  move away from each other and the rogue wave  disappears.  The rogue wave is  short-lived, it exists only for $t$ close to $0$. Notice that the rogue wave appearing at $t=0$ has two tops at points C and D.
 }}}
\end{figure}

 There is a remarkable similarity of the predictions of Figures 19-20 to the Aivasovsky's painting     "Devyatiy Val" .  The column of water  is large and does not break at $t=0$ only for certain values of the angle between the two troughs, the parameters in Figures 19-20 are chosen to produce an appropriate angle. The angle between the troughs shown in the Aivasovsky's painting     "Devyatiy Val"   is very close to the angle between the troughs in Figures 19-20. The rogue wave rises up at the intersection of the two troughs both in Figures 19-20 and in the painting.  Is the similarity merely coincidental or the result of  Aivasovsky's numerous discussions with sailors and sea men who undoubtedly had encountered rogue waves? The answer to this question we will never know.
 \begin{figure}[!h]
 {\includegraphics[scale=1.0]{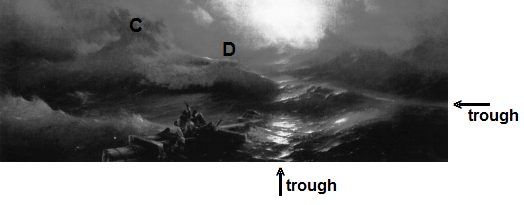}\caption{ Lower half of "Devyatiy Val" by  Aivasovsky. Notice that the  two troughs   intersect at the angle about the same as the angle of intersection of the troughs in Figures 19,20 at $t=0$. The rogue wave in the painting has two tops at points C and D, and so does the rogue wave in Figures 19 and 20 at $t=0$. In the painting the top at D is overturning. The line connecting points C and D is almost parallel to one of the troughs both in Figures 19, 20 and in the painting.  Reproduced from public domain  http://thewallpapers.co.uk/?s=Aivazovsky\%20-\%20The\%20Ninth\%20Wave\&page=1 . }}
\end{figure}

Here we have shortly discussed modeling large waves with the solutions of KP, the topic is elaborated in [Kov 1]. The modeling is more of qualitative than quantitive   nature and is far from perfect. Such models are often called {\it mathematical caricatures} since just like ordinary caricatures they capture only certain important aspects of the original while loosing or distorting others.  Despite its imperfections KP has one advantage over many other models including the nonlinear Schrodinger equation, and that is its two dimensionality as to properly     describe large waves one needs two space dimensions. Models with only one space dimension may be popular and fruitful (in terms of the number of publications one can produce using such models)  but  completely  irrelevant  as far as the physical  waves are concerned.

\section*{   Acknowledgments.  }

The current article is based on [Kov 1, Kov 2] so references are made only to these articles, for further references the reader is referred to bibliography in  [Kov 1, Kov 2].

The author would like to thank Professor Hubert Chanson for kindly providing his picture of an undular bore; Roger Brooke and Andres Grawin  for permission to reproduce the picture of the tsunami of December 26, 2004;  John Wormald for permission to reproduce the picture of the wave at the
Hells Mouth, Wales;  Scott McClimont of Rip Curl and the authors for permission to use the frames from the video "Tip 2 Tip - Seven Ghosts  The Teaser" ;   Dr. Susanne Lehner and    Deutsches Zentrum fur Luft- und Raumfahrt for permission  to reproduce the ESA images.   The original pictures were trimmed and turned into black and white  to minimize the cost of publication  by the author of this paper who takes the responsibility for any loss in technical or artistic quality of the pictures.

   \subsubsection{\rm \bf Bibliography.}

\noindent   [Kov1]   Kovalyov, M., On the nature of large and rogue wave, posted at arXiv, identifier 1208.2047

\noindent   [Kov 2]  Kovalyov, M., On a class of solutions of KdV. { \it  Journal of Differential Equations}, 2005,   vol. 213, issue 1,  pp  1-80.

\noindent [Rip Curl] , video Tip 2 Tip - Seven Ghosts, {http://www.youtube.com/watch?v=2\_vmS-bCoNU}

\end{document}